\documentclass[a4paper,fleqn]{cas-sc}
\usepackage[numbers]{natbib}
\usepackage{algorithm}
\usepackage{algpseudocode}
\usepackage{xcolor}
\definecolor{grayblue}{RGB}{200,215,240}
\def\tsc#1{\csdef{#1}{\textsc{\lowercase{#1}}\xspace}}
\tsc{WGM}
\tsc{QE}
\begin{document}
\let\WriteBookmarks\relax
\def\floatpagepagefraction{1}
\def\textpagefraction{.001}
\let\printorcid\relax 

\shorttitle{}    

\shortauthors{Hanyu Ding et al.}

\title[mode = title]{Keyword Mamba: Spoken Keyword Spotting with State Space Models}

\author[1]{Hanyu Ding}
\author[1]{Wenlong Dong}
\author[1,2,3]{Qirong Mao}
\cormark[1]

\affiliation[1]{organization={School of Computer Science and Communication Engineering, Jiangsu University}, addressline={Zhenjiang}, postcode={212013}, country={China}} 
\affiliation[2]{organization={Jiangsu Engineering Research Center of Big Data Ubiquitous Perception and Intelligent Agricultural Applications}, addressline={Zhenjiang}, postcode={212013}, country={China}} 
\affiliation[3]{organization={Provincial Key Laboratory of Computational Intelligence and New Technologies in Low-Altitude Digital Agriculture}, addressline={Zhenjiang}, postcode={212013}, country={China}} 
\cortext[1]{Corresponding author at: School of Computer Science and Communication Engineering, Jiangsu University, Zhenjiang, 212013, China.}  
\cortext[1]{E-mail addresses: mao\_qr@ujs.edu.cn (Q. Mao).}

\begin{abstract}
Keyword spotting (KWS) is an essential task in speech processing. It is widely used in voice assistants and smart devices. Deep learning models like CNNs, RNNs, and Transformers have performed well in KWS. However, they often struggle to handle long-term patterns and stay efficient at the same time. In this work, we present Keyword Mamba, a new architecture for KWS. It uses a neural state space model (SSM) called Mamba. We apply Mamba along the time axis and also explore how it can replace the self-attention part in Transformer models. We test our model on the Google Speech Commands datasets. The results show that Keyword Mamba reaches strong accuracy with fewer parameters and lower computational cost. To our knowledge, this is the first time a state space model has been used for KWS. These results suggest that Mamba has strong potential in speech-related tasks.
\end{abstract}




\maketitle

\section{Introduction}
Keyword spotting (KWS) \cite{lopez2021deep} refers to the task of detecting specific keywords in audio streams comprising speech. As a critical entry point for human-computer interaction systems, KWS has been extensively applied in voice assistants, smart home devices, and intelligent cockpit systems \cite{ng23b_interspeech,zhang2017hello,dar2025bi,peng2025dark,xiao2025analytickws}. Accurate keyword detection enables these applications to be activated in response to user commands, thereby initiating subsequent operations. Consequently, the advancement of KWS techniques has become a key area of research to enable more seamless human-computer interaction experiences.

\begin{figure}[h]
    \centering
    \includegraphics[width=0.8\textwidth]{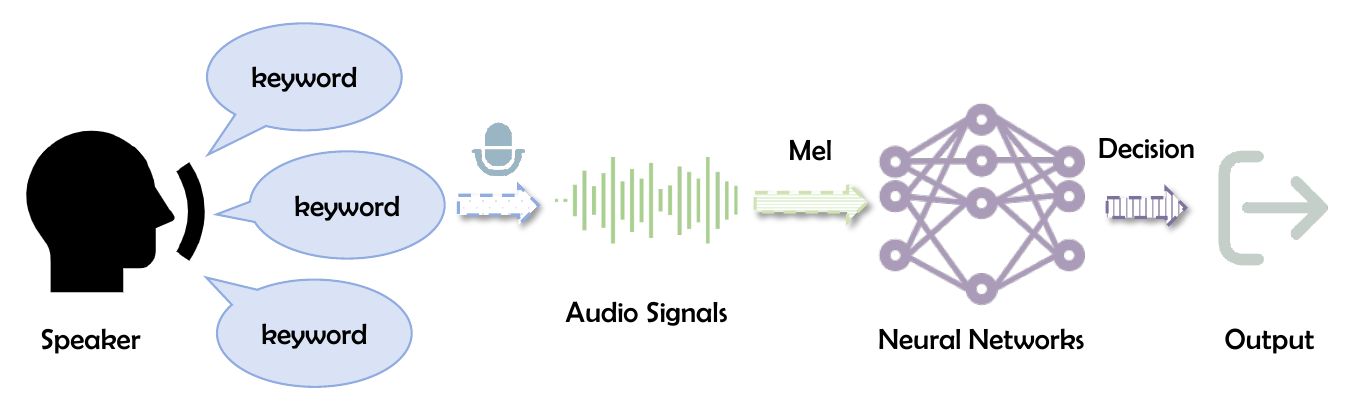}
    \vspace{-2mm}
    \caption{Keyword spotting using deep learning illustration.}
    \vspace{-5mm}
    \label{fig:KWS}
\end{figure}

In recent years, with the rapid advancement of deep learning technologies, KWS has also experienced significant growth. A major milestone in this development was the introduction of the first deep learning-based speech KWS system \cite{chen2014small} in 2014, which marked the beginning of a new era in the field. Fig. \ref{fig:KWS} shows an illustration of the keyword spotting process using deep learning, in which spoken words are converted into audio signals and processed by neural networks. Since then, the field of KWS has witnessed a surge of research adopting representative neural network architectures, with convolutional neural networks (CNNs) \cite{lecun2002gradient,choi19_interspeech,kim21l_interspeech}, recurrent neural networks (RNNs) \cite{de2018neural,hochreiter1997long,rybakov20_interspeech}, and Transformers \cite{vaswani2017attention,berg21_interspeech} emerging as the predominant frameworks. Notably, BC-ResNet \cite{kim21l_interspeech}, a convolutional neural network based on broadcasted residual learning, combines the strengths of 1D temporal and 2D spatial convolutions while minimizing additional computational cost, and demonstrates strong performance in the KWS task. On the RNN side, MHAtt-RNN \cite{rybakov20_interspeech} introduces multi-head attention mechanisms, which significantly improve the model’s ability to capture key temporal dependencies in speech sequences. In addition, Transformer-based architectures such as the Keyword Transformer (KWT) \cite{berg21_interspeech} further advance the field by leveraging self-attention to effectively model long-range contextual information.

Despite the success of these models in KWS applications, several inherent limitations still persist. For instance, the fixed receptive field of CNNs restricts their ability to capture long-range temporal dependencies. RNNs, due to their sequential computation nature, are difficult to parallelize, resulting in slower training and inference speeds. Moreover, Transformer models suffer from quadratic growth in computational complexity with respect to the size of the input context window, which significantly limits their efficiency and scalability when processing long sequences.

To address these inherent challenges, a new architecture based on neural state space models (SSMs) \cite{kalman1960new,gu2022efficiently,gu2022parameterization} known as Mamba \cite{gu2024mamba}, has recently emerged and achieved performance surpassing state-of-the-art methods across a range of tasks \cite{zhang2025rethinking,shi2024multi,pmlr-v235-zhu24f}. In the field of speech processing, recent studies have explored replacing Transformers with Mamba for tasks such as speech enhancement \cite{zhang2025mamba} and automatic speech recognition \cite{zhang2025mamba}, demonstrating promising initial results. Since different speech tasks \cite{xiao2025xlsr,xiao2024tf,jiang2025dual} focus on various characteristics of the speech signal, such as speaker identity, linguistic content, or emotional state, it remains unclear how to effectively apply the Mamba architecture to other speech-related tasks such as spoken KWS.

In this work, inspired by Mamba and Vision Mamba (Vim) \cite{pmlr-v235-zhu24f}, we introduce them to KWS by developing a novel architecture called Keyword Mamba (KWM). Our objective is to explore the application of Mamba in KWS. Specifically, we apply Mamba solely along the temporal domain, leveraging its capability to effectively model long-range dependencies in order to enhance KWS performance while maintaining computational efficiency. In addition, to further investigate the potential of Mamba, we attempt to replace the multi-head self-attention (MHSA) module \cite{vaswani2017attention} in the Transformer architecture with Mamba, rather than replacing the entire model, aiming to improve the nonlinear modeling capacity of Mamba. We evaluate Keyword Mamba on various versions of the Google Speech Commands datasets \cite{warden2018speech}, demonstrating its effectiveness and robustness in real-world scenarios. To the best of our knowledge, this is the first study to apply a state space-based model to KWS, highlighting the novelty and significance of our approach.

\section{Related Works}
\subsection{State Space Models}
SSMs \cite{kalman1960new,gu2022efficiently,gu2022parameterization,gu2024mamba} are statistical frameworks designed for modeling time-series data. They are widely used in machine learning and statistics to handle dynamic systems and time-varying processes. SSMs are capable of modeling the evolution of latent system states over time and capturing the relationships between these hidden states and the observed data. Importantly, SSMs not only support long-range dependency modeling but also offer linear computational complexity. Recently, the advantages of SSMs in sequence analysis have gained more and more attention from scholars. Among them, the structured state space sequence model (S4) \cite{gu2022efficiently} and the selective state space model (Mamba) \cite{gu2024mamba} are typically represented.

Structured State Space Sequence Model (S4). The S4 is an efficient sequence modeling approach designed to capture long-range dependencies using SSMs. It incorporates three core mechanisms: HiPPO \cite{gu2020hippo} for memory of input history, Diagonal Plus Low-Rank Parameterization for stability and diagonalizability, and Efficient Kernel Computation via FFTs and iFFTs. The S4 significantly reduces computational complexity to \(O(N \ log(N))\) and achieves strong performance on tasks like path-X in the LRA benchmark. Moreover, its effectiveness extends to various other benchmark tasks as well. Although this model still falls short of state-of-the-art Transformers in terms of performance, it continues to close the gap and provides improved computational efficiency \cite{fu2023simple}.

Selective State Space Model (Mamba). Based on the S4, Mamba is proposed. Mamba is a more efficient state space modeling approach to address the limitations of traditional SSMs in tasks such as selective copying and induction \cite{olsson2022context}, while avoiding the quadratic computational complexity \(O(N^2)\) and memory complexity of Transformers. It adopts an input-aware parameterization scheme combined with a simplified selection mechanism and introduces an efficient, hardware-friendly selective scan algorithm. Mamba also employs a gating mechanism to reduce the dimensionality of kernel operations and integrates a gated MLP \cite{liu2021pay} to further enhance its modeling capability. In both computer vision (CV) \cite{pmlr-v235-zhu24f} and natural language processing (NLP) \cite{waleffe2406empirical} tasks, Mamba has demonstrated outstanding performance and effectiveness. Meanwhile, with linear-time complexity \(O(N)\), Mamba is more efficient than traditional Transformers.

\subsection{Visual Mamba}
With the ongoing development and adoption of Mamba, the model has been rapidly extended to the visual domain. Vision Mamba (Vim) \cite{pmlr-v235-zhu24f}, one of the earliest and most representative visual Mamba models, draws inspiration from Vision Transformers (ViT) \cite{dosovitskiy2021an} by incorporating position embeddings for spatial information encoding and leveraging bidirectional SSMs to process non-causal image sequences. Another variant, VMamba \cite{liu2024vmamba}, unfolds image patches into sequences along both the horizontal and vertical axes, enabling bidirectional scanning in both directions. Likewise, several other notable works have investigated Mamba-based visual backbones \cite{li2024mamba,pei2025efficientvmamba,zhan2024exploring,xie2024quadmamba}, consistently reporting competitive performance across various vision tasks, including classification, detection, and segmentation.

\subsection{Speech Mamba}
Recent works have explored the use of SSMs and Mamba methods in speech processing tasks. Notably, several studies investigated self-supervised audio Mamba models trained using masked spectrogram modeling. For example, Audio Mamba (AuM) \cite{erol2024audio} and its variant for audio tagging achieve comparable or superior performance to Audio Spectrogram Transformers (AST) \cite{gong2022ssast}, while using only about one-third of the parameters. Furthermore, Mamba has been applied to specific tasks such as speech enhancement \cite{chao2024investigation,quan2024multichannel}, automatic speech recognition \cite{gao2024speech,masuyama2024mamba} and speech separation \cite{jiang2025dual,jiang2025speech,dang2024u}. However, the effective design of Mamba models for deep KWS remains unexplored. Unlike these other tasks, KWS is less sensitive to frequency information \cite{choi19_interspeech}. Therefore, we propose applying Mamba solely within the temporal domain. This leverages its ability to model long-range dependencies efficiently, aiming to enhance KWS performance while maintaining efficiency.

\section{Proposed Keyword Mamba}
The goal of Keyword Mamba is to introduce the advanced SSM (i.e., Mamba) into the KWS task and explore how to effectively apply Mamba in this context. This section begins with the fundamental principles of Mamba, followed by a comprehensive overview of the Keyword Mamba framework and a detailed explanation of its key architectural component (i.e., Mamba Encoder).

\subsection{Preliminaries}
Mamba is built upon the S4, retaining its core SSM framework while introducing a selective mechanism to enhance its ability to focus on task-relevant features.

\subsubsection{Structured State Space Sequence Model (S4)} The core dynamics of SSMs lie in how it updates the latent state from one time step to the next, and how it derives the output sequence from this latent state:

\begin{equation}
\begin{aligned}
&h'(t) = A h(t) + B x(t), \\
&y(t) = C h(t).
\end{aligned}
\end{equation}

Here, $A$ denotes the state transition matrix, $B$ is the input-to-state projection matrix, and $C$ is the state-to-output projection matrix. For the rest, $x(t)$ represents the input at time step 
$t$, $h(t)$ denotes the latent state, and $y(t)$ is the generated output. Since most practical applications involve discrete data such as text, it is necessary to discretize SSMs. This is achieved by introducing a special fourth parameter, $\Delta$, which is used to convert the continuous parameters $A$ and $B$ into their discrete counterparts:

\begin{equation}
\begin{aligned}
&\overline{A} = \exp(\Delta A), \\
&\overline{B} = (\Delta A)^{-1} \left( \exp(\Delta A) - I \right) \cdot \Delta B.
\end{aligned}
\end{equation}

Here, $\overline{A}$ and $\overline{B}$ represent the parameters obtained through the discretization of their continuous-time counterparts. After applying discretization, the equations take the following form:

\begin{equation}
\begin{aligned}
&h'(t) = \overline{A} h(t-1) + \overline{B} x(t), \\
&y(t) = C h(t).
\end{aligned}
\end{equation}

To simplify calculations, the repeated application of the equation can be efficiently performed simultaneously using a global convolution approach:

\begin{equation}
\begin{aligned}
&y = x \circledast \overline{K}, \\
&\text{where } \overline{K} = \left( C\overline{B},\, C\overline{A} \overline{B},\, \ldots,\, C\overline{A}^{L-1} \overline{B} \right).
\end{aligned}
\end{equation}

Here, $L$ is the length of the input sequence $x$, $\circledast$ denotes convolution operation, and $\overline{K}$ is the SSM kernel.

\subsubsection{Selective State Space Model (Mamba)} Based on the S4, Mamba improves the performance of SSMs by introducing Selective State Space Models, allowing the continuous parameters to vary with the input enhances selective information processing across sequences, which extend the discretization process by selection mechanism \cite{han2024demystify}:

\begin{equation}
\begin{aligned}
&\overline{B} = s_B(x), \quad \overline{C} = s_C(x), \\
&\Delta = \tau_A\left(\text{Parameter} + s_A(x)\right).
\end{aligned}
\end{equation}

Here, $s_B(x)$ and $s_C(x)$ are linear functions that project input $x$ into an $N$-dimensional space, while $s_A(x)$ broadens a $D$-dimensional linear projection to the fixed dimensions.

\begin{figure*}[t!]
    \centering
    \includegraphics[width=0.9\textwidth]{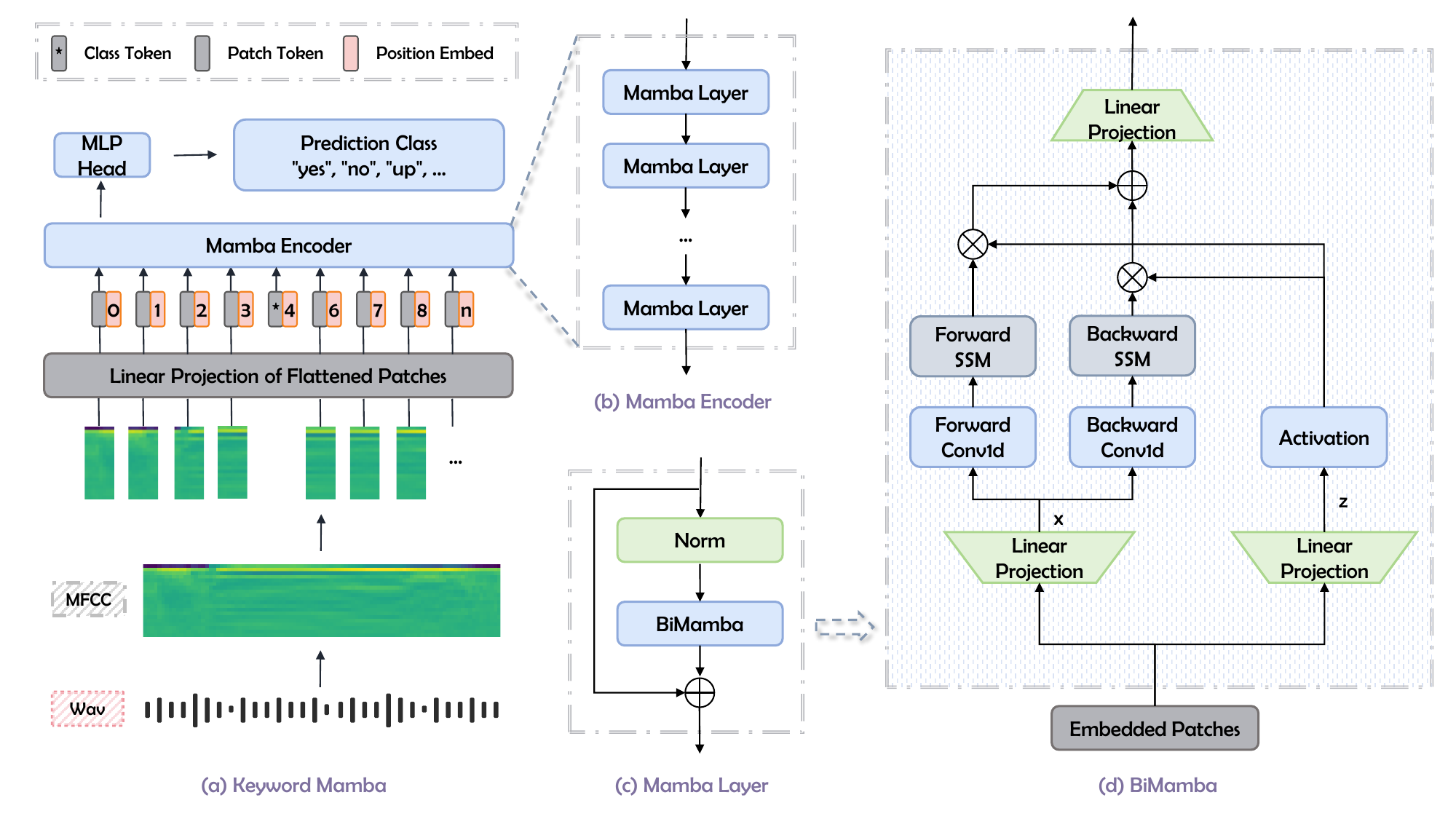}
    \caption{Overview of the Keyword Mamba architecture for keyword spotting: (a) Keyword Mamba; (b) Mamba Encoder of Keyword Mamba; (c) Constituent Layers of Mamba Encoder; (d) Bidirectional Mamba (BiMamba) of Constituent Layers.}
    \label{fig:Keyword Mamba}
    \vspace{-10mm}
\end{figure*}

\subsection{Keyword Mamba}

The Keyword Mamba architecture, as depicted in Fig. \ref{fig:Keyword Mamba}, begins by transforming an input audio waveform into the MFCC  spectrogram \cite{davis1980comparison} \(M \in \mathbb{R}^{F \times T} \), where \(F\) and \(T\) represent the frequency and time dimensions, respectively. The spectrogram is partitioned into a sequence of \(N\) patches \( X \in \mathbb{R}^{N \times (f \cdot t)} \), with \(f\) and \(t\) denoting the shape of each patch and \(N\) calculated as \(N = (F/f) \times  (T/t)\). Since Mamba is applied exclusively in the time domain, \(f = F\), \(t = 1\), and \(N = T\), which means that after flattening, the sequence of patches can be represented as \(X \in \mathbb{R}^{T \times F} \). Then, the sequence is mapped to a higher dimension \( d \) using a linear projection matrix \( W_0 \in \mathbb{R}^{F \times d} \) in the frequency domain. To learn a global representation of the entire spectrogram, we insert a learnable class embedding \cite{dosovitskiy2021an} \( X_{\text{class}} \in \mathbb{R}^{1 \times d} \) into the middle of the projected sequence like Vim \cite{pmlr-v235-zhu24f}. Specifically, the class token is placed between the first and second halves of the sequence after projection. A learnable positional embedding \( X_{\text{pos}} \in \mathbb{R}^{(T+1) \times d} \) is then added, and the final input to the Mamba Encoder is defined as:
\begin{equation}
\begin{aligned}
&X_0 = 
\left[ X_{1:\lfloor T/2 \rfloor} W_0;\ X_{\text{class}};\ 
X_{\lfloor T/2 \rfloor+1:T} W_0 \right] + X_{\text{pos}}.
\end{aligned}
\end{equation}

Then, the resulting sequence \( X_{l-1} \) is fed into the \( l \)-layer of the Mamba Encoder to produce the output \( X_l \). The final output is obtained by normalizing the class token \( X_L^0 \), and then passing it through an MLP head for classification, where 
\( L \) is the number of layers:
\begin{equation}
\begin{aligned}
&X_l = \text{Mamba Encoder}(X_{l-1}) + X_{l-1}, \\
&f = \text{Norm}(X_L^0), \quad p = \text{MLP}(f).
\end{aligned}
\end{equation}

\begin{figure*}[t!]
    \centering
    \includegraphics[width=0.9\textwidth]{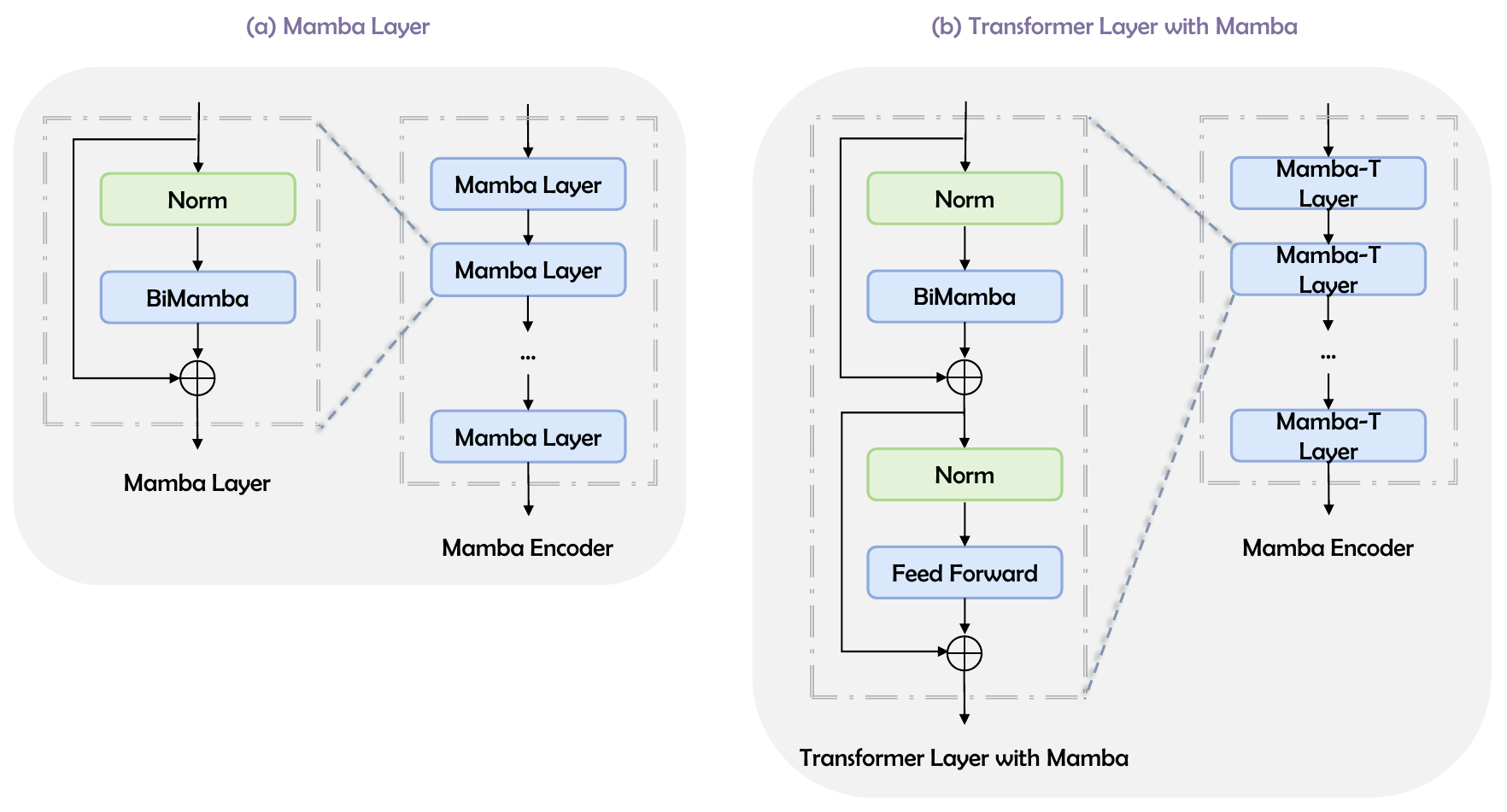}
    \vspace{-1mm}
    \caption{Task-Aware Encoder Designs: (a) Mamba Layer: using stacked BiMamba layers as an alternative to Transformer layers; (b) Transformer Layer with Mamba:  replacing MHSA in Transformer layer with BiMamba.}
    \label{fig:Encoder}
    \vspace{-6mm}
\end{figure*}

\subsection{Mamba Encoder}
The Mamba Encoder serves as the core architectural component of Keyword Mamba. In this section, we first present the specific Mamba structure (i.e. BiMamba) employed in our work, followed by the Mamba-based encoder designed specifically for the KWS task.
\subsubsection{BiMamba}
The original Mamba \cite{gu2024mamba} performs causal computations in a unidirectional manner, relying solely on historical information. However, bidirectional processing is essential for speech tasks in which the entire utterance is available \cite{zhang2025mamba}. To address this, Mamba requires to support the bidirectional processing, similar to the MHSA module, in order to capture global dependencies within the input signal features. In this study, we employ the Bidirectional Mamba design (BiMamba) \cite{pmlr-v235-zhu24f} inspired by Vim, aiming to enhance the model’s ability to capture contextual dependencies in both temporal directions. In other words, BiMamba enables the model to process information from past and future contexts.

The BiMamba block is shown in Fig. \ref{fig:Keyword Mamba}(d). Specifically, BiMamba employs two sets of SSM and Conv1D modules (one forward and one backward) that share the same input and output projection layers. The input embedding, which is first processed by the input projection layer, is then fed separately into a forward SSM and a backward SSM, resulting in two outputs. After that, the outputs are combined and passed through the output projection layer.

\subsubsection{Task-Aware Encoder Designs}
For the Mamba Encoder in Keyword Mamba, we explore two different design variants to capture information of various abstraction levels, as shown in Fig. \ref{fig:Encoder}: one composed entirely of Mamba layers, and the other incorporating additional nonlinear transformations by integrating Mamba components into Transformer layers.

\textbf{Mamba Layer.} As depicted in Fig. \ref{fig:Encoder}(a), the first strategy uses the BiMamba layers independently (i.e., as a direct replacement for the Transformer layer) to construct the Mamba Encoder. The objective is to directly leverage Mamba’s ability to capture long-range dependencies within the input data, thereby enhancing the effectiveness of KWS. This design can better focus on temporal patterns relevant to keyword activation, while maintaining a lightweight architecture suitable for real-time or low-resource scenarios. Algo. \ref{alg:mamba-layer} illustrates the workflow of the Mamba Layer.

\textbf{Transformer Layer with Mamba.} The second approach employs the BiMamba layer to replace the MHSA modules within the Transformer, where the feed-forward network (FFN) and normalization are used to provide additional nonlinearity. This design choice is primarily motivated by the fact that SSMs are largely composed of linear operations. Although the SiLU \cite{elfwing2018sigmoid} activation is used in the residual structure in practice, its capacity to capture high-level information (e.g., semantics) remains limited. Therefore, introducing additional nonlinear components is essential for enhancing Mamba’s ability to model higher-level representations. Algo. \ref{alg:mamba-trans-layer} illustrates the workflow of the Transformer Layer with Mamba.

\section{Experimental Setup}
\subsection{Datasets and Metrics}
We evaluate the performance of the proposed Keyword Mamba model on Google Speech Commands datasets V1 and V2 \cite{warden2018speech}. V1 of the dataset contains 64,727 snippets of 30 different words from various speakers, whereas V2 contains 105,829 snippets of 35 different words. The 12-label classification task uses 10 words: “up”, “down”, “left”, “right”, “yes”, “no”, “on”, “off”, “go”, and “stop”, in addition to “silence” and “unknown”, where instances of the latter is taken from the remaining words in the dataset, whereas the 30-label task and the 35-label task use all available words. We use the same 80:10:10 train/validation/test data split ratio as \cite{rybakov20_interspeech,berg21_interspeech} for side-by-side comparisons and follow the evaluation criteria of \cite{rybakov20_interspeech} as closely as possible.

For these datasets, the primary metric is accuracy (ACC) \cite{rybakov20_interspeech}. For each experiment, we train the model three times with different random initializations. Therefore, the final results are presented as average values.

\begin{algorithm}
\caption{Mamba Layer Workflow}
\label{alg:mamba-layer}
\begin{algorithmic}[1]
\Require $\mathbf{X}_{l-1}: (B, L, D)$
\Ensure $\mathbf{X}_l: (B, L, D)$

\State $\mathbf{X}'_{l-1}: (B, L, D) \gets \textbf{Norm}(\mathbf{X}_{l-1})$
\State $\mathbf{x}: (B, L, E) \gets \textbf{Linear}^\mathbf{x}(\mathbf{X}'_{l-1})$
\State $\mathbf{z}: (B, L, E) \gets \textbf{Linear}^\mathbf{z}(\mathbf{X}'_{l-1})$
\For{$o \in \{\text{forward}, \text{backward}\}$}
    \State $\mathbf{x}'_o: (B, L, E) \gets \textbf{SiLU}(\textbf{Conv1d}_o(\mathbf{x}))$
    \State $\mathbf{B}_o: (B, L, N) \gets \textbf{Linear}_o^\mathbf{B}(\mathbf{x}'_o)$
    \State $\mathbf{C}_o: (B, L, N) \gets \textbf{Linear}_o^\mathbf{C}(\mathbf{x}'_o)$
    \State $\mathbf{\Delta}_o: (B, L, N) \gets \log(1 + \exp(\textbf{Linear}_{o}^{\mathbf{\Delta}}(\mathbf{x}'_o) + \textbf{Parameter}_{o}^{\mathbf{\Delta}}))$
    \State $\overline{\mathbf{A}}_o: (B, L, E, N) \gets \mathbf{\Delta}_o \otimes \textbf{Parameter}_{o}^\mathbf{A}$
    \State $\overline{\mathbf{B}}_o: (B, L, E, N) \gets \mathbf{\Delta}_o \otimes \mathbf{B}_o$
    \State $\mathbf{y}_o: (B, L, E) \gets \textbf{SSM}(\overline{\mathbf{A}}_o, \overline{\mathbf{B}}_o, \mathbf{C}_o)(\mathbf{x}'_o)$
\EndFor
\State $\textbf{y}'_{\text{forward}}: (B, L, E) \gets \textbf{y}_{\text{forward}} \odot \textbf{SiLU}(\mathbf{z})$
\State $\textbf{y}'_{\text{backward}}: (B, L, E) \gets \textbf{y}_{\text{backward}} \odot \textbf{SiLU}(\mathbf{z})$
\State $\mathbf{X}_l: (B, L, D) \gets \textbf{Linear}^{\mathbf{X}}\left( \textbf{y}'_{\text{forward}} + \textbf{y}'_{\text{backward}} \right) + \mathbf{X}_{l-1}$

\State \Return $\mathbf{X}_l$
\end{algorithmic}
\end{algorithm}

\begin{algorithm}
\caption{Transformer Layer with Mamba Workflow}
\label{alg:mamba-trans-layer}
\begin{algorithmic}[1]
\Require $\mathbf{X}_{l-1}: (B, L, D)$
\Ensure $\mathbf{X}_l: (B, L, D)$

\State $\mathbf{X}'_{l-1}: (B, L, D) \gets \textbf{Norm}(\mathbf{X}_{l-1})$
\State $\mathbf{x}: (B, L, E) \gets \textbf{Linear}^\mathbf{x}(\mathbf{X}'_{l-1})$
\State $\mathbf{z}: (B, L, E) \gets \textbf{Linear}^\mathbf{z}(\mathbf{X}'_{l-1})$
\For{$o \in \{\text{forward}, \text{backward}\}$}
    \State $\mathbf{x}'_o: (B, L, E) \gets \textbf{SiLU}(\textbf{Conv1d}_o(\mathbf{x}))$
    \State $\mathbf{B}_o: (B, L, N) \gets \textbf{Linear}_o^\mathbf{B}(\mathbf{x}'_o)$
    \State $\mathbf{C}_o: (B, L, N) \gets \textbf{Linear}_o^\mathbf{C}(\mathbf{x}'_o)$
    \State $\mathbf{\Delta}_o: (B, L, N) \gets \log(1 + \exp(\textbf{Linear}_{o}^{\mathbf{\Delta}}(\mathbf{x}'_o) + \textbf{Parameter}_{o}^{\mathbf{\Delta}}))$
    \State $\overline{\mathbf{A}}_o: (B, L, E, N) \gets \mathbf{\Delta}_o \otimes \textbf{Parameter}_{o}^\mathbf{A}$
    \State $\overline{\mathbf{B}}_o: (B, L, E, N) \gets \mathbf{\Delta}_o \otimes \mathbf{B}_o$
    \State $\mathbf{y}_o: (B, L, E) \gets \textbf{SSM}(\overline{\mathbf{A}}_o, \overline{\mathbf{B}}_o, \mathbf{C}_o)(\mathbf{x}'_o)$
\EndFor
\State $\textbf{y}'_{\text{forward}}: (B, L, E) \gets \textbf{y}_{\text{forward}} \odot \textbf{SiLU}(\mathbf{z})$
\State $\textbf{y}'_{\text{backward}}: (B, L, E) \gets \textbf{y}_{\text{backward}} \odot \textbf{SiLU}(\mathbf{z})$
\State $\tilde{\mathbf{X}}_l: (B, L, D) \gets \textbf{Linear}^{\tilde{\mathbf{X}}}\left( \textbf{y}'_{\text{forward}} + \textbf{y}'_{\text{backward}} \right) + \mathbf{X}_{l-1}$

\State $\tilde{\mathbf{X}}'_{l}: (B, L, D) \gets \textbf{Norm}(\tilde{\mathbf{X}}_{l})$
\State $\mathbf{X}_l: (B, L, D) \gets \textbf{FeedForward}^\textbf{X}(\tilde{\mathbf{X}}'_{l}) + \tilde{\mathbf{X}}_{l}$

\Procedure{FeedForward}{$\mathbf{X}$}
    \State $\mathbf{X}: (B, L, D)$
    \State $\mathbf{H}: (B, L, D_{f}) \gets \textbf{Linear}_1(\mathbf{X})$
    \State $\mathbf{G}: (B, L, D_{f}) \gets \mathbf{GELU}(\mathbf{H})$
    \State $\mathbf{O}: (B, L, D) \gets \textbf{Linear}_2(\mathbf{G})$
    \State \Return $\mathbf{O}$
\EndProcedure

\State \Return $\mathbf{X}_l$
\end{algorithmic}
\end{algorithm}

\subsection{Implementation Details}
The model is trained for 200 epochs on V1 and 140 epochs on V2 with a batch size of 128, using the AdamW optimizer. The initial learning rate is set to 0.001, and a cosine learning rate schedule is applied. Additionally, the training includes 10 warmup epochs. For regularization, a weight decay of 0.1 and label smoothing of 0.1 are used. As for the loss function, the cross-entropy loss is used.

During preprocessing, the input audio is converted into a sequence of 40-dimensional MFCC calculated from 30ms window with 10ms stride. Then, we pad the time dimension with zeros to a fixed length of 98 feature vectors per sample. This means that the frequency dimension of MFCC is 40 and the time dimension is 98. To improve generalization, various data augmentation techniques are applied when training. These include time shifting in the range of [-100, 100] ms, resampling within [0.85, 1.15], and adding background noise with a volume of 0.1. In addition, 2 time masks are applied, with sizes in the range of [0, 25], and 2 frequency masks are applied, with sizes in the range of [0, 7] \cite{park19e_interspeech}. 

\subsection{Model Variants}
Due to the parameter constraints inherent and lightweight requirements in the KWS task, we adopt the same dimensional configurations as KWT \cite{berg21_interspeech}, setting the model dimensions to 192, 128, and 64, respectively.

Since the Mamba Encoder in the Keyword Mamba comprises two distinct variants, for ease of distinction in our subsequent experiments, we denote the inclusion of the Mamba Layer and the Transformer Layer with Mamba as KWM and KWM-T, respectively.

\section{Results and Analysis}
\subsection{Keyword Mamba vs. Other SOTA Models}
To ensure a fair comparison with other state-of-the-art (SOTA) models, we set the number of layers in Keyword Mamba to 12, matching the configuration of the KWT series \cite{berg21_interspeech}. Table \ref{tab:t1} summarizes the average accuracy across four evaluation settings from the Speech Commands V1 and V2 datasets. As shown in the results, Keyword Mamba consistently outperforms existing models across all datasets, establishing new benchmark records. These gains are not only in accuracy but also in parameter count and efficiency.

Relative to Transformer-based models like KWT, Keyword Mamba achieves better accuracy with fewer parameters, demonstrating its efficiency in both training and inference. For instance, while KWT-3 requires over 5M parameters, Keyword Mamba reaches higher accuracy with only 3.4M parameters, confirming its strong parameter efficiency.

In addition, Keyword Mamba shows competitive or superior performance when compared to CNN-based models such as BC-ResNet \cite{kim21l_interspeech} and MatchboxNet \cite{majumdar20_interspeech}. It also surpasses RNN-based models like Att-RNN \cite{de2018neural} and MHAtt-RNN \cite{rybakov20_interspeech}, which typically struggle with long-range temporal patterns and parallel computation. It is worth emphasizing that this is the first known application of Mamba to the keyword spotting task. Despite its novelty, Keyword Mamba not only performs well but also demonstrates strong generalization across different subsets. These results clearly support the effectiveness of state space models in speech tasks, especially in learning discriminative and time-aware features from audio signals.

\begin{table*}[t!]
    \centering
    \caption{Accuracy comparison of various models on Speech Commands V1 and V2 benchmarks. KWM and KWM-T achieve state-of-the-art performance while maintaining smaller model sizes compared to Transformer-based baselines. The results exceeding existing best methods are \colorbox{grayblue}{\textbf{highlighted}}.}
    \vspace{-1mm}
    \label{tab:t1}
    \setlength{\tabcolsep}{15pt}
    \begin{tabular}{llllll}
    \toprule
        \textbf{Model} & \textbf{Para} & \textbf{V1-12} & \textbf{V1-30} & \textbf{V2-12} & \textbf{V2-35} \\
    \midrule
        Att-RNN \cite{de2018neural} & 202K & 95.6 & 96.29 & 96.9 & 93.9  \\ 
        MHAtt-RNN \cite{rybakov20_interspeech} & 743K & 97.2 & 97.02 & 98.0 & 97.27  \\ 
        TC-ResNet14-1.5 \cite{choi19_interspeech} & 305K & 96.6 & 96.99 & 97.43 & 95.43  \\ 
        Matchboxnet-3x2x64 \cite{majumdar20_interspeech} & 93K & 97.48 & 97.23 & 97.21 & 97.46  \\ 
        BC-ResNet-8 \cite{kim21l_interspeech} & 321K & 98.0 & 97.46 & 98.7 & 97.65  \\ 
        ConvMixer \cite{ng2022convmixer} & 119K & 97.3 & 97.05 & 98.2 & 96.83  \\ 
        DenseNet-BiLSTM \cite{zeng2019effective} & 250K & 97.5 & / & 97.4 & /  \\ 
        SincConv+DSConv \cite{mittermaier2020small} & 122K & 96.6 & / & 97.4 & /  \\ 
        SincConv+GDSConv \cite{mittermaier2020small} & 62K & 96.4 & / & 97.3 & /  \\ 
        NoisyDARTS-TC14 \cite{zhang2021autokws} & 108K & 96.79±0.30 & / & 97.18±0.26 & /  \\ 
        LG-Net6 \cite{wang21da_interspeech} & 321K & 97.67 & / & 96.79 & /  \\ 
        KWT-3 \cite{berg21_interspeech} & 5361K & 97.49±0.15 & 96.83 & 98.56±0.07 & 97.69±0.09  \\ 
        KWT-2 \cite{berg21_interspeech} & 2394K & 97.27±0.08 & 96.34 & 98.43±0.08 & 97.74±0.03  \\ 
        KWT-1 \cite{berg21_interspeech} & 607K & 97.26±0.18 & 95.96 & 98.08±0.10 & 96.95±0.14  \\ 
    \midrule
        KWM-192 & 3.4M & \cellcolor{grayblue}{\textbf{98.01}} & \cellcolor{grayblue}{\textbf{97.73}} & \cellcolor{grayblue}{\textbf{98.79}} & \cellcolor{grayblue}{\textbf{97.86}}  \\ 
        KWM-128 & 1.6M & 97.95 &  \cellcolor{grayblue}{\textbf{97.54}} & 98.62 & \cellcolor{grayblue}{\textbf{97.84}}  \\ 
        KWM-64 & 0.5M & 97.46 & 97.07 & 98.13 & 97.12  \\
    \midrule
        KWM-T-192 & 5.2M &  \cellcolor{grayblue}{\textbf{98.05}} &  \cellcolor{grayblue}{\textbf{97.80}} &  \cellcolor{grayblue}{\textbf{98.91}} &  \cellcolor{grayblue}{\textbf{97.89}}  \\
        KWM-T-128 & 2.4M & 97.88 &  \cellcolor{grayblue}{\textbf{97.70}} & \cellcolor{grayblue}{\textbf{98.73}} &  \cellcolor{grayblue}{\textbf{97.86}}  \\
        KWM-T-64 & 0.7M & 97.72 &  \cellcolor{grayblue}{\textbf{97.69}} & 98.56 &  \cellcolor{grayblue}{\textbf{97.75}}  \\ 
    \bottomrule
    \end{tabular}
    \vspace{-4mm}
\end{table*}

\subsection{Keyword Mamba - KWM (Mamba Layer)}
To explore whether the success of Mamba in other domains can be transferred to the KWS task, we conduct detailed experiments on the KWM model using pure Mamba layers. Specifically, we vary both the model dimension (192, 128, 64) and layer depth (12, 10, 8, 6), as shown in Table~\ref{tab:t2}. The results show that KWM maintains high and stable accuracy across different configurations. For example, KWM-192 with 12 layers achieves 98.01\% on V1-12 and 98.79\% on V2-12. Even when the model is reduced to just 6 layers and a dimension of 64, it still reaches 96.74\%–97.45\%, which is comparable to many larger CNN or RNN models. This behavior reveals a key strength of Mamba-based designs: robust performance under compression. Unlike Transformer-based or convolutional models that often lose accuracy when downsized, KWM continues to perform well with significantly fewer parameters. This makes it ideal for real-world applications where computational and memory resources are limited, such as mobile or embedded devices. The strong performance of small KWM models is likely due to Mamba’s ability to model long-range dependencies efficiently without relying on attention mechanisms. Since Mamba uses a recurrent state-based structure, it can capture temporal patterns over extended time spans, even in smaller architectures. This explains why deeper or wider configurations bring improvements, but are not essential for competitive accuracy. In summary, this experiment supports our core claim: KWM is scalable and effective across different model sizes, proving that the state space modeling capability of Mamba translates well to the audio domain, especially for KWS tasks.

\begin{table*}[t!]
    \centering
    \caption{Effect of Mamba layer depth and model size (dimension) on accuracy. KWM shows strong performance even at low parameter counts, confirming its scalability and efficiency. The best results are \colorbox{blue!20}{\textbf{highlighted}}.}
    \vspace{-1mm}
    \label{tab:t2}
    \setlength{\tabcolsep}{26pt}
    \begin{tabular}{llllll}
    \toprule
        \textbf{Model} & \textbf{Para} & \textbf{V1-12} & \textbf{V1-30} & \textbf{V2-12} & \textbf{V2-35} \\
    \midrule
        \textbf{Dim 192} & ~ & ~ & ~ & ~ &   \\
    \midrule
        KWM-12 & 3.4M & \cellcolor{blue!20}{\textbf{98.01}} & \cellcolor{blue!20}{\textbf{97.73}} & \cellcolor{blue!20}{\textbf{98.79}} & \cellcolor{blue!20}{\textbf{97.86}}  \\ 
        KWM-10 & 2.9M & 97.95 & 97.54 & 98.42 & 97.62  \\ 
        KWM-8 & 2.3M & 97.70 & 97.41 & 98.40 & 97.62  \\ 
        KWM-6 & 1.7M & 97.92 & 97.39 & 98.52 & 97.44  \\ 
    \midrule
        \textbf{Dim 128} & ~ & ~ & ~ & ~ &   \\
    \midrule
        KWM-12 & 1.6M &  \cellcolor{blue!20}{\textbf{97.95}} &  \cellcolor{blue!20}{\textbf{97.54}} &  \cellcolor{blue!20}{\textbf{98.62}} &  \cellcolor{blue!20}{\textbf{97.84}}  \\ 
        KWM-10 & 1.4M & 97.62 & 97.44 & 98.48 &  97.67  \\ 
        KWM-8 & 1.1M & 97.75 & 97.29 & 98.52 & 97.45  \\ 
        KWM-6 & 0.8M & 97.33 & 97.11 & 98.31 & 97.32  \\ 
    \midrule
        \textbf{Dim 64} & ~ & ~ & ~ & ~ &   \\ 
    \midrule
        KWM-12 & 0.5M &  \cellcolor{blue!20}{\textbf{97.46}} &  \cellcolor{blue!20}{\textbf{97.07}} &  \cellcolor{blue!20}{\textbf{98.13}} &  \cellcolor{blue!20}{\textbf{97.12}}  \\ 
        KWM-10 & 0.4M & 97.27 & 96.98 & 97.92 & 97.09  \\ 
        KWM-8 & 0.3M &  97.43 & 96.71 & 98.01 & 97.00  \\ 
        KWM-6 & 0.2M & 96.74 & 96.52 & 97.45 & 96.92  \\ 
    \bottomrule
    \end{tabular}
    \vspace{-5mm}
\end{table*}

\subsection{Keyword Mamba - KWM-T (Transformer Layer with Mamba)}

To further explore how Mamba can enhance KWS, we construct a hybrid model called KWM-T. In this variant, we keep the feed-forward layer from the Transformer structure but replace the multi-head self-attention module with Mamba, as shown in Table~\ref{tab:t3}. This design aims to combine Mamba’s efficient sequence modeling with the Transformer’s strong nonlinear transformation capacity.

We replicate the model size and depth settings used in Section 5.2 to allow direct comparison. As reported in Table~\ref{tab:t3}, KWM-T consistently achieves competitive or even higher accuracy than the original KWM under similar parameter budgets. For example, KWM-T-192 reaches 98.05\% on V1-12 and 98.91\% on V2-12, surpassing both KWM-192 and most Transformer baselines from Table~\ref{tab:t1}.

Even under reduced settings, such as KWM-T-64, the model achieves 97.72\% on V1-12 and 97.75\% on V2-35, confirming that this hybrid design maintains strong performance across a wide range of resource levels. This result supports a key finding: Integrating Mamba with Transformer-style feed-forward layers leads to better KWS performance by combining efficient memory modeling with strong nonlinear capacity.

The feed-forward layer helps Mamba go beyond linear state-space behavior, allowing it to better represent complex speech features. Meanwhile, Mamba replaces the attention mechanism, offering lower computational cost and better scalability for longer input sequences. In summary, KWM-T offers a highly effective trade-off between accuracy and complexity, and the results confirm that this hybrid architecture is a promising design choice for future speech models.

\begin{table*}[t!]
    \centering
    \caption{Effect of Mamba-integrated Transformer depth and size on accuracy. KWM-T models benefit from replacing self-attention with Mamba, achieving high accuracy with moderate complexity.}
    \label{tab:t3}
    \setlength{\tabcolsep}{25.5pt}
    \begin{tabular}{llllll}
    \toprule
        \textbf{Model} & \textbf{Para} & \textbf{V1-12} & \textbf{V1-30} & \textbf{V2-12} & \textbf{V2-35} \\
    \midrule
        \textbf{Dim 192} & ~ & ~ & ~ & ~ &   \\ 
    \midrule
        KWM-T-12 & 5.2M & \cellcolor{blue!20}{\textbf{98.05}} &  97.80 &  \cellcolor{blue!20}{\textbf{98.91}} &  \cellcolor{blue!20}{\textbf{97.89}}  \\ 
        KWM-T-10 & 4.3M & 97.85 & \cellcolor{blue!20}{\textbf{97.86}} & 98.62 & 97.85  \\ 
        KWM-T-8 & 3.5M & 97.85 & 97.69 & 98.58 & 97.67  \\ 
        KWM-T-6 & 2.6M & 97.62 & 97.57 & 98.42 & 97.63  \\
    \midrule
        \textbf{Dim 128} & ~ & ~ & ~ & ~ &   \\
    \midrule
        KWM-T-12 & 2.4M & \cellcolor{blue!20}{\textbf{97.88}} & \cellcolor{blue!20}{\textbf{97.70}} & \cellcolor{blue!20}{\textbf{98.73}} & \cellcolor{blue!20}{\textbf{97.86}}  \\ 
        KWM-T-10 & 2.0M & 97.69 & 97.55 & 98.52 & 97.76  \\ 
        KWM-T-8 & 1.6M & 97.49 & 97.64 & 98.48 & 97.64  \\ 
        KWM-T-6 & 1.2M & 97.56 & 97.57 & 98.36 & 97.49  \\
    \midrule
       \textbf{Dim 64} & ~ & ~ & ~ & ~ &   \\
    \midrule
        KWM-T-12 & 0.7M & \cellcolor{blue!20}{\textbf{97.72}} & \cellcolor{blue!20}{\textbf{97.69}} & \cellcolor{blue!20}{\textbf{98.56}} & \cellcolor{blue!20}{\textbf{97.75}}  \\ 
        KWM-T-10 & 0.6M & 97.49 & 97.38 & 98.33 & 97.61  \\ 
        KWM-T-8 & 0.5M & 97.40 & 97.23 & 98.09 & 97.45  \\ 
        KWM-T-6 & 0.4M & 97.49 & 97.00 & 98.17 & 97.22  \\ 
    \bottomrule
    \end{tabular}
\end{table*}

\begin{figure}[t!]
    \centering
    \includegraphics[width=0.6\textwidth]{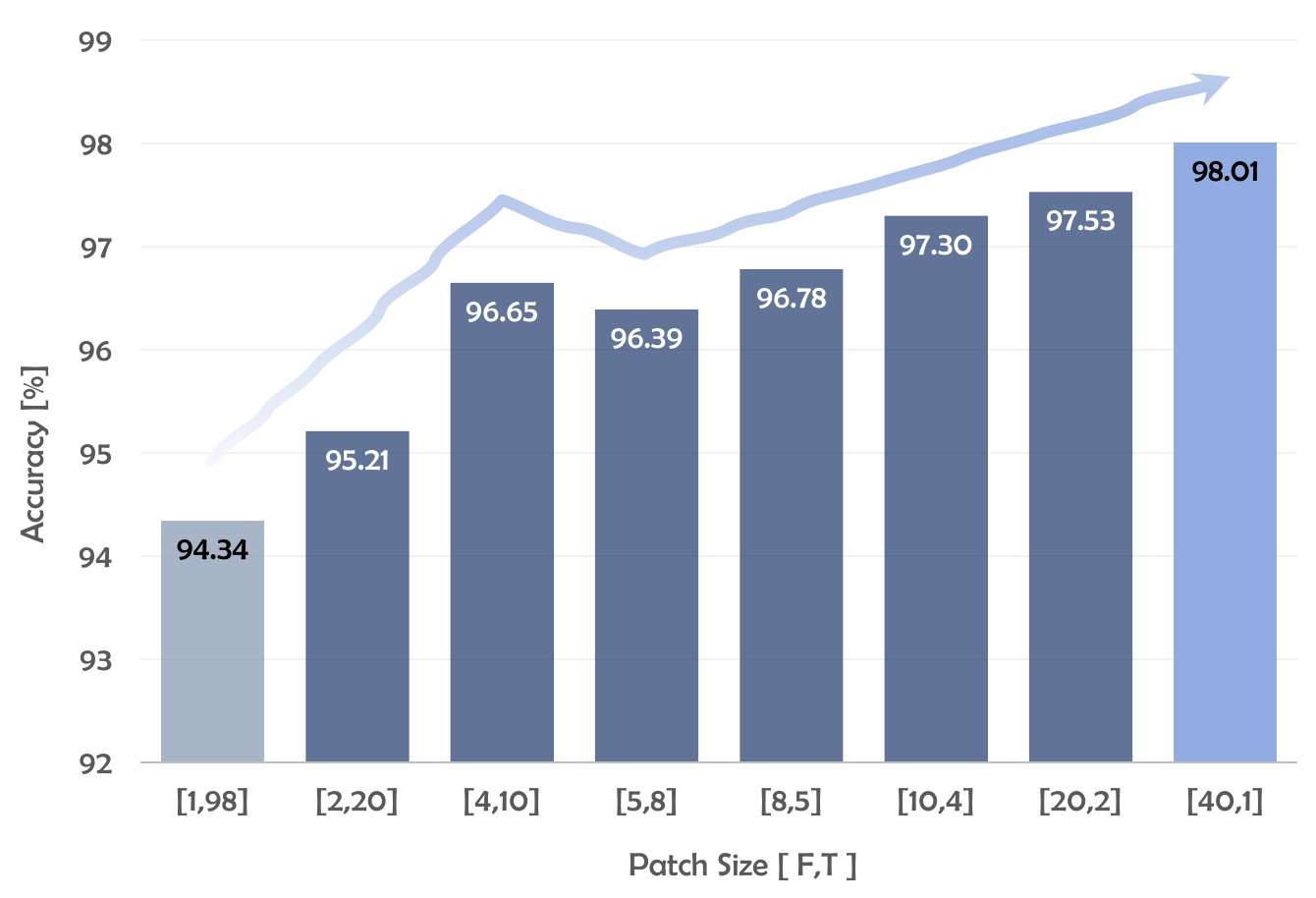}
    \caption{Impact of patch size on keyword spotting accuracy. Larger temporal patches significantly improve performance, suggesting that long-range temporal modeling is critical. Results are based on the Speech Commands V1-12 dataset using KWM-192.}
    \label{fig:Patch}
    \vspace{-4mm}
\end{figure}

\subsection{Ablation Studies}

We conduct three ablation studies to understand how different design choices affect the performance of Keyword Mamba. These studies focus on (1) the shape of MFCC patches, (2) the position of the class token, and (3) the directionality of the Mamba components.

\textbf{Temporal vs. Frequency Modeling:} First, we vary the shape of MFCC spectrogram patches fed into the model. Our baseline uses temporal-domain Mamba with patches shaped [40,1], but we also test frequency-domain Mamba and rectangular patches that mix both time and frequency dimensions.

\begin{table}[t!]
    \centering
    \setlength{\tabcolsep}{13pt}
    \caption{Ablation study on class token position and Mamba type. Mid-position tokens and bidirectional Mamba configurations (BiMamba-Bi-Bi) provide the best accuracy, highlighting the importance of token placement and directional flow.}
    \label{tab:t4}
    \begin{tabular}{lll}
    \toprule
        \textbf{Ablation Objective} & \textbf{Type} & \textbf{V1-12} \\
    \midrule
    \multirow{3}{*}{class token position} 
    & Mid & \cellcolor{blue!20}{\textbf{98.01}} \\
    & Head & 97.75 \\
    & End & 97.92 \\
    \midrule
    \multirow{3}{*}{mamba type}
    & BiMamba-Bi-Bi & \cellcolor{blue!20}{\textbf{98.01}} \\
    & BiMamba-Fo-Bi &  97.69\\
    & Mamba-Fo-Fo &  78.29\\
    \bottomrule
    \end{tabular}
    \vspace{-4mm}
\end{table}

As shown in Fig. \ref{fig:Patch}, the temporal configuration achieves the highest accuracy, reaching 98.01\%. This confirms that modeling along the time axis is most effective for keyword spotting. Our first projection layer (e.g., Patching) acts like a temporal convolution with a kernel size [40,1] and stride 1, which supports earlier findings that temporal convolutions are especially useful for speech recognition~\cite{choi19_interspeech}.

\textbf{Class Token Position:}
Next, we study the effect of class token placement. We test inserting the token at the start, middle, or end of the sequence. Table~\ref{tab:t4} shows that placing the token in the middle of the input achieves the best result (98.01\%), likely because it allows the token to receive information from both earlier and later parts of the sequence.

However, we also find that the class token position has a relatively small impact on performance in keyword spotting. This contrasts with other speech or language tasks, where token position often plays a larger role. 

\textbf{Directional Flow in Mamba Variants:} 
Finally, we compare three types of Mamba configurations based on their directional flow in both the SSM module and Conv1D layers:

\begin{itemize}
    \item BiMamba-Bi-Bi: bidirectional SSM and Conv1D
    \item BiMamba-Fo-Bi: only bidirectional SSM
    \item Mamba-Fo-Fo: unidirectional in both modules
\end{itemize}

Among the three, BiMamba-Bi-Bi achieves the best performance (98.01\%), as shown in Table~\ref{tab:t4}. In contrast, Mamba-Fo-Fo drops significantly to 78.29\%, showing that unidirectional modeling lacks important context. These results confirm the importance of bidirectional information flow in speech tasks, where keyword boundaries and surrounding sounds may appear in both past and future contexts \cite{zhang2025mamba}. Our findings reinforce the idea that capturing both directions is essential for high-performance speech modeling.

\section{Conclusions}
In this work, we propose Keyword Mamba, a novel architecture that brings the Mamba framework into the KWS domain. By applying Mamba along the temporal axis, we effectively capture long-range dependencies while maintaining computational efficiency. To further explore its additional nonlinear potential, we integrate Mamba into the Transformer architecture by replacing only the multi-head self-attention (MHSA) module. Experiments we conducted on multiple versions of the Google Speech Commands datasets demonstrate the effectiveness and robustness of our approach. These results highlight the promising potential of Mamba-based architectures in outperforming traditional Transformers for the KWS task. The implementation
of this work is available at the following GitHub repository: https://github.com/dhyzy123/KWM.

\section*{CRediT authorship contribution statement}
\textbf{Hanyu Ding:} Writing - original draft, Writing - review \& editing, Methodology, Investigation, Validation, Data curation, Software. \textbf{Wenlong Dong:} Writing - review \& editing, Supervision, Formal analysis. \textbf{Qirong Mao:} Writing - review \& editing, Resources, Funding acquisition.

\section*{Declaration of competing interest}
The authors declare that they have no known competing financial interests or personal relationships that could have appeared to influence the work reported in this paper.

\section*{Acknowledgements}
This research was supported by the National Natural Science Foundation of China (No.62176106), the Special Scientific Research Project of the School of Emergency Management of Jiangsu University (No.KY-A-01), the Project of Faculty of Agricultural Engineering of Jiangsu University (No.NGXB20240101), the Key Project of National Natural Science Foundation of China (No.U1836220), and the Jiangsu Key Research and Development Plan Industry Foresight and Key Core Technology (No.BE2020036).

\section*{Data availability}
Data will be made available on request.

\bibliographystyle{cas-model2-names}
\bibliography{kwm-ref}



\end{document}